\documentclass{article}

\usepackage{arxiv}

\usepackage[utf8]{inputenc} 
\usepackage[T1]{fontenc}    
\usepackage{hyperref}       
\usepackage{url}            
\usepackage{booktabs}       
\usepackage{amsfonts}       
\usepackage{amssymb}        
\usepackage{amsmath}        
\usepackage{nicefrac}       
\usepackage{microtype}      
\usepackage{lipsum}		
\usepackage{graphicx}
\usepackage[numbers]{natbib}
\usepackage{doi}
\usepackage{wrapfig}

\title{Stokes's theorem in R}

\author{ \href{https://orcid.org/0000-0001-5982-0415}{\includegraphics[width=0.03\textwidth]{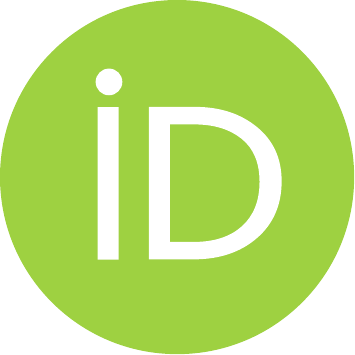}\hspace{1mm}Robin K. S.~Hankin}\thanks{\href{https://academics.aut.ac.nz/robin.hankin}{work};  
\href{https://www.youtube.com/watch?v=JzCX3FqDIOc&list=PL9_n3Tqzq9iWtgD8POJFdnVUCZ_zw6OiB&ab_channel=TrinTragulaGeneralRelativity}{play}} \\
 Auckland University of Technology\\
	\texttt{hankin.robin@gmail.com} \\
}

\renewcommand{\shorttitle}{Stokes's theorem in R}

\hypersetup{
pdftitle={Stokes's theorem in R},
pdfsubject={q-bio.NC, q-bio.QM},
pdfauthor={Robin K. S.~Hankin},
pdfkeywords={Stokes's theorem, exterior calculus, differential forms}
}

\usepackage{Sweave}
\begin{document}
\maketitle

\begin{abstract}
  In this short article I introduce the {\tt stokes} package which
  provides functionality for working with tensors, alternating forms,
  wedge products, and related concepts from the exterior calculus.
  Notation and spirit follow Spivak.  Stokes's generalized integral
  theorem, viz $\int_{\partial X}\phi=\int_Xd\phi$, is demonstrated
  here using the package; it is available on CRAN at\\
  \url{https://CRAN.R-project.org/package=stokes}.
\end{abstract}

\section{Introduction}

\setlength{\intextsep}{0pt}
\begin{wrapfigure}{r}{0.2\textwidth}
  \begin{center}
\includegraphics[width=1in]{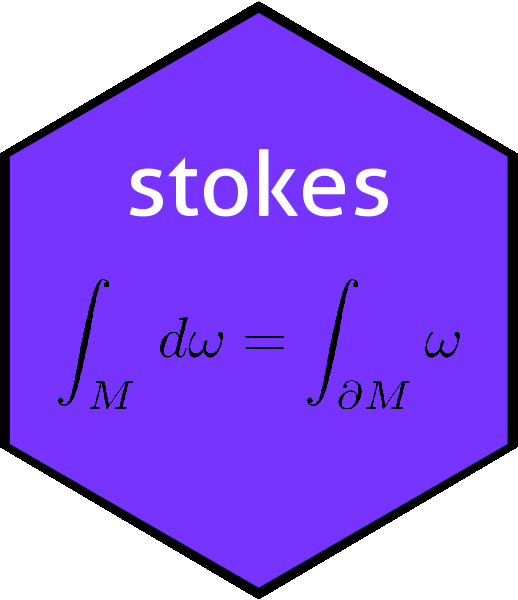}
  \end{center}
\end{wrapfigure}
Ordinary differential calculus may be formalized and generalized to
arbitrary-dimensional oriented manifolds using the exterior calculus.
Here I introduce the {\tt stokes} package, written in the R computing
language~\citep{rcore2022}, which furnishes functionality for working
with the exterior calculus.  I give numerical verification of a number
of theorems using the package.  Notation follows that of
Spivak~\cite{spivak1965}.

\section{Tensors}

Recall that a $k$-tensor is a multilinear map $S\colon
V^k\longrightarrow\mathbb{R}$, where $V=\mathbb{R}^n$ is considered as
a vector space; Spivak denotes the space of multilinear maps as
$\mathcal{J}^k(V)$.  Formally, multilinearity means

\begin{equation}
S\left(v_1,\ldots,av_i,\ldots,v_k\right)=a\cdot S\left(v_1,\ldots,v_i,\ldots,v_k\right)
\end{equation}

and

\begin{equation}
S\left(v_1,\ldots,v_i+{v_i}',\ldots,v_k\right)=S\left(v_1,\ldots,v_i,\ldots,x_v\right)+
S\left(v_1,\ldots,{v_i}',\ldots,v_k\right).
\end{equation}

where $v_i\in V$ and $a\in\mathbb{R}$.  If $S\in\mathcal{J}^k(V)$ and
$T\in\mathcal{J}^l(V)$, then we may define $S\otimes
T\in\mathcal{J}^{k+l}(V)$ as

\begin{equation}
S\otimes T\left(v_1,\ldots,v_k,v_{k+1},\ldots,v_{k+l}\right)=
S\left(v_1,\ldots,v_k\right)\cdot
T\left(v_1,\ldots,v_l\right)
\end{equation}

Spivak observes that $\mathcal{J}^k(V)$ is spanned by the $n^k$
products of the form

\begin{equation}
\phi_{i_1}\otimes\phi_{i_2}\otimes\cdots\otimes\phi_{i_k}\qquad
1\leq i_i,i_2,\ldots,i_k\leq n
\end{equation}

where $v_1,\ldots,v_k$ is a basis for $V$ and
$\phi_i\left(v_j\right)=\delta_{ij}$; we can therefore write

\begin{equation}
S=\sum_{1\leq i_1,\ldots,i_k\leq n} a_{i_1\ldots i_k}
\phi_{i_1}\otimes\cdots\otimes\phi_{i_k}.
\end{equation}

The space spanned by such products has a natural representation in R
as an array of dimensions $n\times\cdots\times n=n^k$.  If {\tt A} is
such an array, then the element {\tt A[i\_1,i\_2,...,i\_k]} is the
coefficient of $\phi_{i_1}\otimes\ldots\otimes\phi_{i\_k}$.  However,
it is more efficient and conceptually cleaner to consider a {\em
sparse} array, as implemented by the {\tt spray}
package~\citep{hankin2022_spray}.  We will consider the case
$n=5,k=4$, so we have multilinear maps from
$\left(\mathbb{R}^5\right)^4$ to $\mathbb{R}$.  Below, we will test
algebraic identities in R using the idiom furnished by the {\tt
stokes} package.  For our example we will define
$S=1.5\phi_5\otimes\phi_1\otimes\phi_1\otimes\phi_1+2.5\phi_1\otimes\phi_1\otimes\phi_2\otimes\phi_3+3.5\phi_1\otimes\phi_3\otimes\phi_4\otimes\phi_2$
using a matrix with three rows, one per term, and whose rows
correspond to each term's tensor products of the $\phi$'s.  We first
have to load the {\tt stokes} package:

\begin{Schunk}
\begin{Sinput}
> library("stokes")
\end{Sinput}
\end{Schunk}

Then the idiom is straightforward:

\begin{Schunk}
\begin{Sinput}
> k <- 4
> n <- 5
> M <- matrix(c(5,1,1,1, 1,1,2,3, 1,3,4,2),3,4,byrow=TRUE)
> M
\end{Sinput}
\begin{Soutput}
     [,1] [,2] [,3] [,4]
[1,]    5    1    1    1
[2,]    1    1    2    3
[3,]    1    3    4    2
\end{Soutput}
\begin{Sinput}
> S <- as.ktensor(M,coeffs= 0.5 + 1:3)
> S
\end{Sinput}
\begin{Soutput}
A linear map from V^4 to R with V=R^5:
             val
 5 1 1 1  =  1.5
 1 1 2 3  =  2.5
 1 3 4 2  =  3.5
\end{Soutput}
\end{Schunk}

Observe that, if stored as an array of size $n^k$, $S$ would have
$5^4=625$ elements, all but three of which are zero.  So $S$ is a
4-tensor, mapping $V^4$ to $\mathbb{R}$, where $V=\mathbb{R}^5$.  Here
we have
$S=1.5\phi_5\otimes\phi_1\otimes\phi_1\otimes\phi_1+2.5\phi_1\otimes\phi_1\otimes\phi_2\otimes\phi_3+3.5\phi_1\otimes\phi_3\otimes\phi_4\otimes\phi_2$.
Note that in some implementations the row order of object {\tt S} will
differ from that of {\tt M}; this phenomenon is due to the underlying
C implementation using the {\tt STL map} class and is discussed in
more detail in the {\tt spray} and {\tt disordR}
packages~\cite{hankin2022_disordR,hankin2022_spray}.

\subsection{Package idiom for evaluation of a tensor}

First, we will define $E$ to be a random point in $V^k$ in terms of a
matrix:

\begin{Schunk}
\begin{Sinput}
> set.seed(0)
> (E <- matrix(rnorm(n*k),n,k))   # A random point in V^k
\end{Sinput}
\begin{Soutput}
           [,1]         [,2]       [,3]       [,4]
[1,]  1.2629543 -1.539950042  0.7635935 -0.4115108
[2,] -0.3262334 -0.928567035 -0.7990092  0.2522234
[3,]  1.3297993 -0.294720447 -1.1476570 -0.8919211
[4,]  1.2724293 -0.005767173 -0.2894616  0.4356833
[5,]  0.4146414  2.404653389 -0.2992151 -1.2375384
\end{Soutput}
\end{Schunk}

Recall that $n=5$, $k=4$, so $E\in\left(\mathbb{R}^5\right)^4$.  We
can evaluate $S$ at $E$ as follows:

\begin{Schunk}
\begin{Sinput}
> f <- as.function(S)
> f(E)
\end{Sinput}
\begin{Soutput}
[1] -3.068997
\end{Soutput}
\end{Schunk}

\subsection{Vector space structure of tensors}

Tensors have a natural vector space structure; they may be added and
subtracted, and multiplied by a scalar, the same as any other vector
space.  Below, we define a new tensor $S_1$ and work with $2S-3S_1$:

\begin{Schunk}
\begin{Sinput}
> S1 <- as.ktensor(1+diag(4),1:4)
> 2*S-3*S1
\end{Sinput}
\begin{Soutput}
A linear map from V^4 to R with V=R^5:
             val
 5 1 1 1  =    3
 1 2 1 1  =   -6
 1 1 2 3  =    5
 1 3 4 2  =    7
 1 1 1 2  =  -12
 1 1 2 1  =   -9
 2 1 1 1  =   -3
\end{Soutput}
\end{Schunk}

We may verify that tensors are linear using package idiom:

\begin{Schunk}
\begin{Sinput}
> LHS <- as.function(2*S-3*S1)(E)
> RHS <- 2*as.function(S)(E) -3*as.function(S1)(E)
> c(lhs=LHS,rhs=RHS,diff=LHS-RHS)
\end{Sinput}
\begin{Soutput}
          lhs           rhs          diff 
 2.374816e+00  2.374816e+00 -4.440892e-16 
\end{Soutput}
\end{Schunk}

(that is, identical up to numerical precision).

\subsection{Numerical verification of multilinearity in the package}

Testing multilinearity is straightforward in the package.  To do this,
we need to define three matrices {\tt E1,E2,E3} corresponding to
points in $\left(\mathbb{R}^5\right)^4$ which are identical except for
one column.  In {\tt E3}, this column is a linear combination of the
corresponding column in {\tt E2} and {\tt E3}:

\begin{Schunk}
\begin{Sinput}
> E1 <- E
> E2 <- E
> E3 <- E
> x1 <- rnorm(n)
> x2 <- rnorm(n)
> r1 <- rnorm(1)
> r2 <- rnorm(1)
> E1[,2] <- x1
> E2[,2] <- x2
> E3[,2] <- r1*x1 + r2*x2
\end{Sinput}
\end{Schunk}

Then we can verify the multilinearity of $S$ by coercing to a function
which is applied to {\tt E1, E2, E3}:

\begin{Schunk}
\begin{Sinput}
> f <- as.function(S)
> LHS <- r1*f(E1) + r2*f(E2)
> RHS <- f(E3)
> c(lhs=LHS,rhs=RHS,diff=LHS-RHS)
\end{Sinput}
\begin{Soutput}
       lhs        rhs       diff 
-0.5640577 -0.5640577  0.0000000 
\end{Soutput}
\end{Schunk}

(that is, identical up to numerical precision).  Note that this is
{\em not} equivalent to linearity over $V^{nk}$:

\begin{Schunk}
\begin{Sinput}
> E1 <- matrix(rnorm(n*k),n,k)
> E2 <- matrix(rnorm(n*k),n,k)
> LHS <- f(r1*E1+r2*E2)
> RHS <- r1*f(E1)+r2*f(E2)
> c(lhs=LHS,rhs=RHS,diff=LHS-RHS)
\end{Sinput}
\begin{Soutput}
       lhs        rhs       diff 
 0.1731245  0.3074186 -0.1342941 
\end{Soutput}
\end{Schunk}

\subsection{Tensor product of general tensors}

Given two k-tensor objects $S,T$ we can form the tensor product
$S\otimes T$, defined as

\begin{equation}\label{definition_of_tensor_cross_product}
S\otimes T\left(v_1,\ldots,v_k,v_{k+1},\ldots, v_{k+l}\right)=
  S\left(v_1,\ldots v_k\right)\cdot T\left(v_{k+1},\ldots
  v_{k+l}\right)
\end{equation}

We will calculate the tensor product of two tensors {\tt S1,S2}
defined as follows:

\begin{Schunk}
\begin{Sinput}
> (S1 <- ktensor(spray(cbind(1:3,2:4),1:3)))
\end{Sinput}
\begin{Soutput}
A linear map from V^2 to R with V=R^4:
         val
 1 2  =    1
 2 3  =    2
 3 4  =    3
\end{Soutput}
\begin{Sinput}
> (S2 <- as.ktensor(matrix(1:6,2,3)))
\end{Sinput}
\begin{Soutput}
A linear map from V^3 to R with V=R^6:
           val
 1 3 5  =    1
 2 4 6  =    1
\end{Soutput}
\end{Schunk}

The R idiom for $S1\otimes S2$ as per
equation~\ref{definition_of_tensor_cross_product} would be {\tt
tensorprod()}, or {\tt \%X\%}:

\begin{Schunk}
\begin{Sinput}
> tensorprod(S1,S2)
\end{Sinput}
\begin{Soutput}
A linear map from V^5 to R with V=R^6:
               val
 1 2 1 3 5  =    1
 3 4 1 3 5  =    3
 1 2 2 4 6  =    1
 3 4 2 4 6  =    3
 2 3 2 4 6  =    2
 2 3 1 3 5  =    2
\end{Soutput}
\end{Schunk}

Then, for example:

\begin{Schunk}
\begin{Sinput}
> E <- matrix(rnorm(30),6,5)
> LHS <- as.function(tensorprod(S1,S2))(E)
> RHS <- as.function(S1)(E[,1:2]) * as.function(S2)(E[,3:5])
> c(lhs=LHS,rhs=RHS,diff=LHS-RHS)
\end{Sinput}
\begin{Soutput}
      lhs       rhs      diff 
-1.048329 -1.048329  0.000000 
\end{Soutput}
\end{Schunk}

(that is, identical up to numerical precision).

\section{Alternating forms}

An alternating form is a multilinear map $T$ satisfying

\begin{equation}\label{definition_of_alternating}
\mathrm{T}\left(v_1,\ldots,v_i,\ldots,v_j,\ldots,v_k\right)=
    -\mathrm{T}\left(v_1,\ldots,v_j,\ldots,v_i,\ldots,v_k\right)
\end{equation}

(or, equivalently,
$\mathrm{T}\left(v_1,\ldots,v_i,\ldots,v_i,\ldots,v_k\right)= 0$).  We
write $\Lambda^k(V)$ for the space of all alternating multilinear maps
from $V^k$ to $\mathbb{R}$.  Spivak gives
$\operatorname{Alt}\colon\mathcal{J}^k(V)\longrightarrow\Lambda^k(V)$
defined by

\begin{equation}\operatorname{Alt}(T)\left(v_1,\ldots,v_k\right)=
    \frac{1}{k!}\sum_{\sigma\in S_k}\operatorname{sgn}(\sigma)\cdot
    T\left(v_{\sigma(1)},\ldots,v_{\sigma(k)}\right)
\end{equation}

where the sum ranges over all permutations of
$\left[n\right]=\left\{1,2,\ldots,n\right\}$ and
$\operatorname{sgn}(\sigma)\in\pm 1$ is the sign of the permutation.
If $T\in\mathcal{J}^k(V)$ and $\omega\in\Lambda^k(V)$, it is
straightforward to prove that $\operatorname{Alt}(T)\in\Lambda^k(V)$,
$\operatorname{Alt}\left(\operatorname{Alt}\left(T\right)\right)=\operatorname{Alt}\left(T\right)$,
and $\operatorname{Alt}\left(\omega\right)=\omega$.  In the stokes
package, this is effected by the {\tt Alt()} function:

\begin{Schunk}
\begin{Sinput}
> S1
\end{Sinput}
\begin{Soutput}
A linear map from V^2 to R with V=R^4:
         val
 1 2  =    1
 2 3  =    2
 3 4  =    3
\end{Soutput}
\begin{Sinput}
> Alt(S1)
\end{Sinput}
\begin{Soutput}
A linear map from V^2 to R with V=R^4:
          val
 1 2  =   0.5
 2 1  =  -0.5
 3 4  =   1.5
 2 3  =   1.0
 3 2  =  -1.0
 4 3  =  -1.5
\end{Soutput}
\end{Schunk}

Verifying that {\tt Alt(S1)} is in fact alternating is
straightforward, essentially by directly evaluating
Equation~\ref{definition_of_alternating}:

\begin{Schunk}
\begin{Sinput}
> E <- matrix(rnorm(8),4,2)
> Erev <- E[,2:1]
> as.function(Alt(S1))(E) + as.function(Alt(S1))(Erev)  # should be zero
\end{Sinput}
\begin{Soutput}
[1] 0
\end{Soutput}
\end{Schunk}

However, we can see that this form for alternating tensors (here
called $k$-forms) is inefficient and highly redundant: in this example
there is a {\tt 1 2} term and a {\tt 2 1} term (the coefficients are
equal and opposite).  In this example we have $k=2$ but in general
there would be potentially $k!$ essentially repeated terms which
collectively require only a single coefficient.  The package provides
{\tt kform} objects which are inherently alternating using a more
efficient representation; they are described using wedge products
which are discussed next.

\subsection{Wedge products and the exterior calculus}

This section follows the exposition of~\cite{hubbard2015}, who
introduce the exterior calculus starting with a discussion of
elementary forms, which are alternating forms with a particularly
simple structure.  An example of an elementary
form would be $dx_1\wedge dx_3$ [treated as an indivisible entity],
which is an alternating multilinear map from
$\mathbb{R}^n\times\mathbb{R}^n$ to $\mathbb{R}$ with

\begin{equation}
\left(
dx_1\wedge dx_3
\right)\left(
\begin{pmatrix}a_1\\a_2\\a_3\\ \vdots\\ a_n\end{pmatrix},
\begin{pmatrix}b_1\\b_3\\b_3\\ \vdots\\ b_n\end{pmatrix}
\right)=\mathrm{det}
\begin{pmatrix} a_1 & b_1 \\ a_3 & b_3\end{pmatrix}
=a_1b_3-a_3b_1
\end{equation}

That this is alternating follows from the properties of the
determinant.  In general of course, $dx_i\wedge dx_j\left(
\begin{pmatrix}a_1\\ \vdots\\ a_n\end{pmatrix}, \begin{pmatrix}b_1\\
\vdots\\ b_n\end{pmatrix} \right)=\mathrm{det} \begin{pmatrix} a_i &
b_i \\ a_j & b_j\end{pmatrix}$.  Because such objects are linear, it
is possible to consider sums of elementary forms, such as $dx_1\wedge
dx_2 + 3 dx_2\wedge dx_3$ with

\begin{equation}
\left(
dx_1\wedge dx_2 + 3dx_2\wedge dx_3
\right)\left(
\begin{pmatrix}a_1\\a_2\\ \vdots\\ a_n\end{pmatrix},
\begin{pmatrix}b_1\\b_2\\ \vdots\\ b_n\end{pmatrix}
\right)=\mathrm{det}
\begin{pmatrix} a_1 & b_1\\ a_2 & b_2\end{pmatrix}
+3\mathrm{det}
\begin{pmatrix} a_2 & b_2\\ a_3 & b_3\end{pmatrix}
\end{equation}

or even $K=dx_1\wedge dx_2\wedge dx_3 +5dx_1\wedge dx_2\wedge dx_4$
which would be a linear map from $\left(\mathbb{R}^n\right)^3$ to
$\mathbb{R}$ with

\begin{equation}
\left(
dx_4\wedge dx_2\wedge dx_3 +5dx_1\wedge dx_2\wedge dx_4
\right)\left(
\begin{pmatrix}a_1\\a_2\\ \vdots\\ a_n\end{pmatrix},
\begin{pmatrix}b_1\\b_2\\ \vdots\\ b_n\end{pmatrix},
\begin{pmatrix}c_1\\c_2\\ \vdots\\ c_n\end{pmatrix}
\right)=\mathrm{det}
\begin{pmatrix}
	a_4 & b_4 & c_4\\
	a_2 & b_2 & c_2\\
	a_3 & b_3 & c_3
\end{pmatrix}
+5\mathrm{det}
\begin{pmatrix}
	a_1 & b_1 & c_1\\
	a_2 & b_2 & c_2\\
	a_4 & b_4 & c_4
\end{pmatrix}.
\end{equation}

Defining $K$ has ready R idiom in which we define a matrix whose rows
correspond to the differentials in each term:

\begin{Schunk}
\begin{Sinput}
> M <- matrix(c(4,2,3,1,4,2),2,3,byrow=TRUE)
> M
\end{Sinput}
\begin{Soutput}
     [,1] [,2] [,3]
[1,]    4    2    3
[2,]    1    4    2
\end{Soutput}
\begin{Sinput}
> K <- as.kform(M,c(1,5))
> K
\end{Sinput}
\begin{Soutput}
An alternating linear map from V^3 to R with V=R^4:
           val
 2 3 4  =    1
 1 2 4  =   -5
\end{Soutput}
\end{Schunk}

Function {\tt as.kform()} takes each row of {\tt M} and places the
elements in increasing order; the coefficient will change sign if the
permutation is odd.  Note that the order of the rows in {\tt K} is
immaterial and indeed in some implementations will appear in a
different order: the stokes package uses the {\tt spray}
package~\citep{hankin2022_spray} which in turn utilises the {\tt STL}
map class of {\tt C++}; the corresponding R idiom conforms to {\tt
disordR} discipline~\citep{hankin2022_disordR}.

\subsection{Formal definition of $dx$}

In the previous section we defined objects such as ``$dx_1\wedge
dx_6$" as a single entity.  Here I define the elementary form $dx_i$
formally and in the next section discuss the wedge product $\wedge$.
The elementary form $dx_i$ is simply a map from $\mathbb{R}^n$ to
$\mathbb{R}$ with $dx_i\left(x_1,x_2,\ldots,x_n\right)=x_i$.  Observe
that $dx_i$ is an alternating form, even though we cannot swap
arguments (because there is only one).

\subsubsection{Package idiom for creation of differential forms}

Package idiom for creating an elementary differential form appears
somewhat cryptic at first sight, but is consistent (it is easier to
understand package idiom for creating more complicated alternating
forms, as in the next section).  Suppose we wish to work with $dx_3$:

\begin{Schunk}
\begin{Sinput}
> dx3 <- as.kform(matrix(3,1,1),1)
> options(kform_symbolic_print = NULL) # revert to default print method
> dx3
\end{Sinput}
\begin{Soutput}
An alternating linear map from V^1 to R with V=R^3:
       val
 3  =    1
\end{Soutput}
\end{Schunk}

Interpretation of the output above is not obvious (it is easier to
understand the output from more complicated alternating forms, as in
the next section), but for the moment observe that $dx_3$ is indeed an
alternating form, mapping $\mathbb{R}^n$ to $\mathbb{R}$ with
$dx_3\left(x_1,x_2,\ldots,x_n\right)=x_3$.  Thus, for example:

\begin{Schunk}
\begin{Sinput}
> as.function(dx3)(c(14,15,16))
\end{Sinput}
\begin{Soutput}
[1] 16
\end{Soutput}
\begin{Sinput}
> as.function(dx3)(c(14,15,16,17,18))  # idiom can deal with arbitrary vectors
\end{Sinput}
\begin{Soutput}
[1] 16
\end{Soutput}
\end{Schunk}

and we see that $dx_3$ picks out the third element of a vector.  These
are linear in the sense that we may add and subtract these elementary
forms:

\begin{Schunk}
\begin{Sinput}
> dx5 <- as.kform(matrix(5,1,1),1)
> as.function(dx3 + 2*dx5)(1:10)  # picks out element 3 + 2*element 5
\end{Sinput}
\begin{Soutput}
[1] 13
\end{Soutput}
\end{Schunk}

\subsection{Formal definition of wedge product}

The wedge product maps two alternating forms to another alternating
form; formally we write
$\wedge\colon\Lambda^k(V)\times\Lambda^l(V)\longrightarrow\Lambda^{k+l}(V)$.
Given $\omega\in\Lambda^k(V)$ and $\eta\in\Lambda^l(V)$, Spivak
defines the wedge product $\omega\wedge\eta\in\Lambda^{k+l}(V)$ as

\begin{equation}\label{definition_of_wedge_product}
\omega\wedge\eta={k+l\choose k\quad l}\operatorname{Alt}(\omega\otimes\eta)
\end{equation}

The package includes an extensive numerically-oriented discussion of
Equation~\ref{definition_of_wedge_product} and its implementation in
the {\tt wedge} vignette.

\subsubsection{Evaluation of the wedge product using package idiom}

Wedge products are implemented in the package.  To illustrate this we
define two $k$-forms, {\tt K1} and {\tt K2}:

\begin{Schunk}
\begin{Sinput}
> (K1 <- as.kform(matrix(c(3,5,4, 4,6,1),2,3,byrow=TRUE),c(2,7)))
\end{Sinput}
\begin{Soutput}
An alternating linear map from V^3 to R with V=R^6:
           val
 3 4 5  =   -2
 1 4 6  =    7
\end{Soutput}
\begin{Sinput}
> (K2 <- as.kform(cbind(1:5,3:7),1:5))
\end{Sinput}
\begin{Soutput}
An alternating linear map from V^2 to R with V=R^7:
         val
 1 3  =    1
 5 7  =    5
 2 4  =    2
 4 6  =    4
 3 5  =    3
\end{Soutput}
\end{Schunk}

In symbolic notation, {\tt K1} is equal to $7dx_1\wedge dx_4\wedge
dx_6 -2dx_3\wedge dx_4\wedge dx_5$. and {\tt K2} is $dx_1\wedge dx_3+
2dx_2\wedge dx_4+ 3dx_3\wedge dx_5+ 4dx_4\wedge dx_6+ 5dx_5\wedge
dx_7$.  Package idiom for wedge products is straightforward; the caret
(``\verb|^|") is overloaded to return the wedge product:

\begin{Schunk}
\begin{Sinput}
> K1 ^ K2
\end{Sinput}
\begin{Soutput}
An alternating linear map from V^5 to R with V=R^7:
               val
 1 4 5 6 7  =  -35
 1 3 4 5 6  =  -21
\end{Soutput}
\end{Schunk}

(we might write the product as $-35dx_1\wedge dx_4\wedge dx_5\wedge
dx_6\wedge dx_7 -21dx_1\wedge dx_3\wedge dx_4\wedge dx_5\wedge dx_6$).
See how the wedge product eliminates rows with repeated entries,
gathers permuted rows together (respecting the sign of the
permutation), and expresses the result in terms of elementary forms.
The product is a linear combination of two elementary forms; note that
only two coefficients out of a possible ${7\choose 5}=21$ are nonzero.
Note again that the order of the rows in the product is arbitrary, as
per {\tt disordR} discipline.

\subsubsection{Associativity of the wedge product}

The wedge product has formal properties such as distributivity but by
far the most interesting one is associativity, which I will
demonstrate below:

\begin{Schunk}
\begin{Sinput}
> F1 <- as.kform(matrix(c(3,4,5, 4,6,1,3,2,1),3,3,byrow=TRUE))
> F2 <- as.kform(cbind(1:6,3:8),1:6)
> F3 <- kform_general(1:8,2)
> (F1 ^ F2) ^ F3
\end{Sinput}
\begin{Soutput}
An alternating linear map from V^7 to R with V=R^8:
                   val
 1 2 3 4 5 7 8  =   -5
 1 3 4 5 6 7 8  =   -2
 1 2 3 5 6 7 8  =   11
 1 2 3 4 5 6 8  =    1
 2 3 4 5 6 7 8  =    6
 1 2 3 4 6 7 8  =    2
 1 2 3 4 5 6 7  =    1
 1 2 4 5 6 7 8  =   -5
\end{Soutput}
\begin{Sinput}
> F1 ^ (F2 ^ F3)
\end{Sinput}
\begin{Soutput}
An alternating linear map from V^7 to R with V=R^8:
                   val
 1 2 3 4 5 6 7  =    1
 1 3 4 5 6 7 8  =   -2
 1 2 3 4 5 7 8  =   -5
 1 2 3 4 6 7 8  =    2
 1 2 3 4 5 6 8  =    1
 1 2 3 5 6 7 8  =   11
 2 3 4 5 6 7 8  =    6
 1 2 4 5 6 7 8  =   -5
\end{Soutput}
\end{Schunk}

Note carefully in the above that the terms in \verb|(F1 ^ F2) ^ F3|
and \verb|F1 ^ (F2 ^ F3)| appear in a different order.  They are
nevertheless algebraically identical, as we may demonstrate using the
(overloaded) {\tt ==} operator:

\begin{Schunk}
\begin{Sinput}
> (F1 ^ F2) ^ F3 - F1 ^ (F2 ^ F3)
\end{Sinput}
\begin{Soutput}
The zero alternating linear map from V^7 to R with V=R^n:
empty sparse array with 7 columns
\end{Soutput}
\end{Schunk}

Above we see that the two forms are identical.

\subsection{Multilinearity of $k$-forms}

Spivak observes that $\Lambda^k(V)$ is spanned by
the $n\choose k$ wedge products of the form

\begin{equation}
dx_{i_1}\wedge dx_{i_2}\wedge\ldots\wedge dx_{i_k}\qquad
1\leq i_i<i_2<\cdots <i_k\leq n
\end{equation}

where these products are the elementary forms (compare
$\mathcal{J}^k(V)$, which is spanned by $n^k$ elementary forms).
Formally, multilinearity means every element of the space
$\Lambda^k(V)$ is a linear combination of elementary forms, as
illustrated in the package by function {\tt kform\_general()}.  Consider
the following idiom:

\begin{Schunk}
\begin{Sinput}
> Krel <- kform_general(4,2,1:6)
> Krel
\end{Sinput}
\begin{Soutput}
An alternating linear map from V^2 to R with V=R^4:
         val
 1 2  =    1
 1 3  =    2
 2 3  =    3
 1 4  =    4
 2 4  =    5
 3 4  =    6
\end{Soutput}
\end{Schunk}

Object {\tt Krel} is a two-form, specifically a map from
$\left(\mathbb{R}^4\right)^2$ to $\mathbb{R}$.  Observe that
{\tt Krel} has ${4\choose 2}=6$ components, which do not appear in any
particular order.  Addition of such $k$-forms is straightforward in R
idiom but algebraically nontrivial:

\begin{Schunk}
\begin{Sinput}
> (K1 <- as.kform(matrix(1:4,2,2),c(1,109)))
\end{Sinput}
\begin{Soutput}
An alternating linear map from V^2 to R with V=R^4:
         val
 1 3  =    1
 2 4  =  109
\end{Soutput}
\begin{Sinput}
> (K2 <- as.kform(matrix(c(1,3,7,8,2,4),ncol=2,byrow=TRUE),c(-1,5,4)))
\end{Sinput}
\begin{Soutput}
An alternating linear map from V^2 to R with V=R^8:
         val
 1 3  =   -1
 7 8  =    5
 2 4  =    4
\end{Soutput}
\begin{Sinput}
> K1+K2
\end{Sinput}
\begin{Soutput}
An alternating linear map from V^2 to R with V=R^8:
         val
 2 4  =  113
 7 8  =    5
\end{Soutput}
\end{Schunk}

Above, note how the $dx_2\wedge dx_4$ terms combine [to give {\tt 2 4
    = 113}] and the $dx_1\wedge dx_3$ term vanishes by cancellation.

\subsection{Print methods}

Although the spray form used above is probably the most direct and natural
representation of differential forms in  numerical work, sometimes we need
a more algebraic print method.

\begin{Schunk}
\begin{Sinput}
> U <- ktensor(spray(cbind(1:4,2:5),1:4))
> U
\end{Sinput}
\begin{Soutput}
A linear map from V^2 to R with V=R^5:
         val
 1 2  =    1
 2 3  =    2
 3 4  =    3
 4 5  =    4
\end{Soutput}
\end{Schunk}

we can represent this more algebraically using the {\tt as.symbolic()}
function:

\begin{Schunk}
\begin{Sinput}
> as.symbolic(U)
\end{Sinput}
\begin{Soutput}
[1]  + a*b +2 b*c +3 c*d +4 d*e
\end{Soutput}
\end{Schunk}

In the above, {\tt U} is a multilinear map from
$\left(\mathbb{R}^5\right)^2$ to $\mathbb{R}$.  Symbolically, {\tt a}
represents the map that takes $(a,b,c,d,e)$ to $a$, {\tt b} the map that
takes $(a,b,c,d,e)$ to {\tt b}, and so on.  The asterisk {\tt *} represents
the tensor product $\otimes$.  Alternating forms work similarly but
$k$-forms have different defaults:

\begin{Schunk}
\begin{Sinput}
> K <- kform_general(3,2,1:3)
> K
\end{Sinput}
\begin{Soutput}
An alternating linear map from V^2 to R with V=R^3:
         val
 1 2  =    1
 1 3  =    2
 2 3  =    3
\end{Soutput}
\begin{Sinput}
> as.symbolic(K,d="d",symbols=letters[23:26])
\end{Sinput}
\begin{Soutput}
[1]  + dw^dx +2 dw^dy +3 dx^dy
\end{Soutput}
\end{Schunk}

Note that the wedge product $\wedge$, although implemented in package
idiom as \verb|^| or \verb|%^%|, appears in the symbolic
representation as an ascii caret, \verb|^|.

We can alter the default print method with the {\tt kform\_symbolic\_print}
option, which uses {\tt as.symbolic()}:

\begin{Schunk}
\begin{Sinput}
> options(kform_symbolic_print = "d")
> K
\end{Sinput}
\begin{Soutput}
An alternating linear map from V^2 to R with V=R^3:
 + dx1^dx2 +2 dx1^dx3 +3 dx2^dx3 
\end{Soutput}
\end{Schunk}

This print option works nicely with the {\tt d()} function for elementary
forms:

\begin{Schunk}
\begin{Sinput}
> (d(1) + d(5)) ^ (d(3)-5*d(2)) ^ d(7)
\end{Sinput}
\begin{Soutput}
An alternating linear map from V^3 to R with V=R^7:
 + dx1^dx3^dx7 - dx3^dx5^dx7 -5 dx1^dx2^dx7 +5 dx2^dx5^dx7 
\end{Soutput}
\begin{Sinput}
> options(kform_symbolic_print = NULL) # restore default
\end{Sinput}
\end{Schunk}

\subsection{Contractions}

Given a $k$-form $\phi\colon V^k\longrightarrow\mathbb{R}$ and a
vector $\mathbf{v}\in V$, the {\em contraction} $\phi_\mathbf{v}$ of
$\phi$ and $\mathbf{v}$ is a $k-1$-form with

\begin{equation}
  \phi_\mathbf{v}\left(\mathbf{v}^1,\ldots,\mathbf{v}^{k-1}\right) =
  \phi\left(\mathbf{v},\mathbf{v}^1,\ldots,\mathbf{v}^{k-1}\right)
\end{equation}

if $k>1$; we specify $\phi_\mathbf{v}=\phi(\mathbf{v})$ if $k=1$.
Verification is straightforward:

\begin{Schunk}
\begin{Sinput}
> (o <- rform())  # a random 3-form
\end{Sinput}
\begin{Soutput}
An alternating linear map from V^3 to R with V=R^7:
           val
 5 6 7  =    4
 1 3 7  =    7
 2 3 7  =   -2
 1 5 7  =  -12
 1 2 4  =    1
 4 6 7  =    5
 2 4 6  =   -6
 1 4 6  =    8
\end{Soutput}
\begin{Sinput}
> V <- matrix(runif(21),ncol=3)
> LHS <- as.function(o)(V)
> RHS <- as.function(contract(o,V[,1]))(V[,-1])
> c(LHS=LHS,RHS=RHS,diff=LHS-RHS)
\end{Sinput}
\begin{Soutput}
          LHS           RHS          diff 
 4.512547e-01  4.512547e-01 -4.440892e-16 
\end{Soutput}
\end{Schunk}

It is possible to iterate the contraction process; if we pass a matrix
$V$ to {\tt contract()} then this is interpreted as repeated
contraction with the columns of $V$:

\begin{Schunk}
\begin{Sinput}
> as.function(contract(o,V[,1:2]))(V[,-(1:2),drop=FALSE])
\end{Sinput}
\begin{Soutput}
[1] 0.4512547
\end{Soutput}
\end{Schunk}

If we pass three columns to {\tt contract()} the result is a $0$-form:

\begin{Schunk}
\begin{Sinput}
> contract(o,V)
\end{Sinput}
\begin{Soutput}
[1] 0.4512547
\end{Soutput}
\end{Schunk}

In the above, the result is coerced to a scalar; in order to work with
a formal $0$-form (which is represented in the package as a {\tt spray}
with a zero-column index matrix) we can use the {\tt lose=FALSE} argument:

\begin{Schunk}
\begin{Sinput}
> contract(o,V,lose=FALSE)
\end{Sinput}
\begin{Soutput}
An alternating linear map from V^0 to R with V=R^0:
           val
  =  0.4512547
\end{Soutput}
\end{Schunk}

\subsection{Transformations and pullback}

Suppose we are given a two-form $\omega=\sum_{i<j}a_{ij}dx_i\wedge
dx_j$ and relationships $dx_i=\sum_rM_{ir}dy_r$, then we would have

\begin{equation}
\omega =
    \sum_{i<j}
    a_{ij}\left(\sum_rM_{ir}dy_r\right)\wedge\left(\sum_rM_{jr}dy_r\right).
\end{equation}

  The general situation would be  a $k$-form where we would have

\begin{equation}
\omega=\sum_{i_1<\cdots<i_k}a_{i_1\ldots i_k}dx_{i_1}\wedge\cdots\wedge dx_{i_k}
\end{equation}

giving

\begin{equation}\omega =
    \sum_{i_1<\cdots <i_k}\left[
    a_{i_1<\cdots < i_k}\left(\sum_rM_{i_1r}dy_r\right)\wedge\cdots\wedge\left(\sum_rM_{i_kr}dy_r\right)\right].
\end{equation}

So $\omega$ was given in terms of $dx_1,\ldots,dx_k$ and we have
expressed it in terms of $dy_1,\ldots,dy_k$.  So for example if

\begin{equation}
\omega=
  dx_1\wedge dx_2 + 5dx_1\wedge dx_3\end{equation}

and

\begin{equation}
  \left(
  \begin{array}{l}
  dx_1\\
  dx_2\\
  dx_3
  \end{array}
  \right)=
\left(
\begin{array}{ccc}
1 & 4 & 7\\
2 & 5 & 8\\
3 & 6 & 9\\
\end{array}
\right)  \left(
  \begin{array}{l}
  dy_1\\ dy_2\\   dy_3
  \end{array}
  \right)
\end{equation}

then

\begin{equation}
\begin{array}{ccl}
  \omega &=&
\left(1dy_1+4dy_2+7dy_3\right)\wedge
\left(2dy_1+5dy_2+8dy_3\right)+
5\left(1dy_1+4dy_2+7dy_3\right)\wedge
\left(3dy_1+6dy_2+9dy_3\right)
\\
&=&2dy_1\wedge dy_1+5dy_1\wedge dy_2+\cdots+
5\cdot 7\cdot 6dx_3\wedge dx_2+
5\cdot 7\cdot 9dx_3\wedge dx_3+\\
&=& -33dy_1\wedge dy_2-66dy_1\wedge dy_3-33dy_2\wedge dy_3
\end{array}
\end{equation}

Function {\tt pullback()} does all this:

\begin{Schunk}
\begin{Sinput}
> options(kform_symbolic_print = "dx")   # uses dx etc in print method
> pullback(dx^dy+5*dx^dz, matrix(1:9,3,3))
\end{Sinput}
\begin{Soutput}
An alternating linear map from V^2 to R with V=R^3:
 -33 dx^dy -66 dx^dz -33 dy^dz 
\end{Soutput}
\begin{Sinput}
> options(kform_symbolic_print = NULL) # revert to default
\end{Sinput}
\end{Schunk}

However, it is slow and I am not 100\% sure that there isn't a much
more efficient way to do such a transformation.  There are a few tests
in {\tt tests/testthat}.  Here I show that transformations may be inverted
using matrix inverses:

\begin{Schunk}
\begin{Sinput}
> (o <- 2 * as.kform(2) ^ as.kform(4) ^ as.kform(5))
\end{Sinput}
\begin{Soutput}
An alternating linear map from V^3 to R with V=R^5:
           val
 2 4 5  =    2
\end{Soutput}
\begin{Sinput}
> M <- matrix(rnorm(25),5,5)
\end{Sinput}
\end{Schunk}

Then we will transform according to matrix {\tt M} and then transform
according to the matrix inverse; the functionality works nicely with
magrittr pipes:

\begin{Schunk}
\begin{Sinput}
> o |> pullback(M) |> pullback(solve(M))
\end{Sinput}
\begin{Soutput}
An alternating linear map from V^3 to R with V=R^5:
           val
 3 4 5  =    0
 1 3 4  =    0
 1 2 4  =    0
 1 4 5  =    0
 2 3 4  =    0
 2 4 5  =    2
 1 3 5  =    0
 2 3 5  =    0
 1 2 5  =    0
\end{Soutput}
\end{Schunk}

Above we see many rows with values small enough for the print method
to print an exact zero, but not sufficiently small to be eliminated by
the {\tt spray} internals.  We can remove the small entries with {\tt zap()}:

\begin{Schunk}
\begin{Sinput}
> o |> pullback(M) |> pullback(solve(M)) |> zap()
\end{Sinput}
\begin{Soutput}
An alternating linear map from V^3 to R with V=R^5:
           val
 2 4 5  =    2
\end{Soutput}
\end{Schunk}

See how the result is equal to the original $k$-form $2dy_2\wedge
dy_4\wedge dy_5$.

\subsection{Exterior derivatives}

Given a $k$-form $\omega$, Spivak defines the differential of $\omega$
to be a $(k+1)$-form $d\omega$ as follows.  If

\begin{equation}
\omega =
\sum_{
i_1 < i_2 <\cdots<i_k}
\omega_{i_1i_2\ldots i_k}
dx^{i_1}\wedge
dx^{i_2}\wedge\cdots\wedge dx^{i_k}
\end{equation}

then

\begin{equation}
d\omega =
\sum_{
i_1 < i_2 <\cdots<i_k}
\sum_{\alpha=1}^n D_\alpha\left(\omega_{i_1i_2\ldots i_k}\right)
\cdot
dx^{i_1}\wedge
dx^{i_2}\wedge\cdots\wedge dx^{i_k}
\end{equation}

where $D_if(a)=\lim_{h\longrightarrow
0}\frac{f(a^1,\ldots,a^i+h,\ldots,a^n)-f(a^1,\ldots,a^i,\ldots,a^n)}{h}$
is the ordinary $i^\mathrm{th}$ partial derivative (Spivak, p25).
This definition allows one to express the fundamental theorem of
calculus in an arbitrary number of dimensions without modification.
If $f\colon\mathbb{R}^n\longrightarrow\mathbb{R}$ is a scalar function
of position, it can be shown that

\begin{equation}\label{definition_of_d}
    {d}\left(f\,dx_{i_1}\wedge\cdots\wedge dx_{i_k}\right)=
    {d}f\wedge dx_{i_1}\wedge\cdots\wedge dx_{i_k}
\end{equation}

The package provides {\tt grad()} which, when given a vector
$x_1,\ldots,x_n$ returns the one-form

\begin{equation}
\sum_{i=1}^n x_idx_i
\end{equation}

This is useful because $df=\sum_{j=1}^n\left(D_j f\right)\,dx_j$; we
see that $df$ is a one-form that corresponds to the gradient $\nabla
f$ of $f$ in elementary calculus: given a vector $u$ at point $p$ the
correspondence would be $\left(\nabla f\right)\cdot u=
df\left(u\right)$.  Thus

\begin{Schunk}
\begin{Sinput}
> grad(c(0.4,0.1,-3.2,1.5))
\end{Sinput}
\begin{Soutput}
An alternating linear map from V^1 to R with V=R^4:
        val
 1  =   0.4
 2  =   0.1
 3  =  -3.2
 4  =   1.5
\end{Soutput}
\end{Schunk}

We will use the {\tt grad()} function to verify that, in
$\mathbb{R}^n$, a certain $(k-1)$-form has zero work function.
Following Hubbard and Hubbard~\cite{hubbard2015}, we observe that

\begin{equation}
F_3=\frac{1}{\left(x^2+y^2+z^2\right)^{3/2}}
\begin{pmatrix}x\\y\\z\end{pmatrix}
\end{equation}

is a divergenceless velocity field in $\mathbb{R}^3$, and thus motivated
define

\begin{equation}
\omega_n=
d\frac{1}{\left(x_1^2+\ldots +x_n^2\right)^{n/2}}\sum_{i=1}^{n}(-1)^{i-1}
x_idx_1\wedge\cdots\wedge\widehat{dx_i}\wedge\cdots\wedge dx_n
\end{equation}

(where a hat indicates the absence of a term).  Hubbard and Hubbard
show analytically that $d\omega=0$.  Here I verify this reasoning
numerically, using package idiom.  First we define a function that
implements the wedge product
$dx_1\wedge\cdots\wedge\widehat{dx_i}\wedge\cdots\wedge dx_n$:

\begin{Schunk}
\begin{Sinput}
> hat <- function(x){
+     n <- length(x)
+     as.kform(t(apply(diag(n)<1,2,which)))
+ }
\end{Sinput}
\end{Schunk}

So, for example:

\begin{Schunk}
\begin{Sinput}
> hat(1:5)
\end{Sinput}
\begin{Soutput}
An alternating linear map from V^4 to R with V=R^5:
             val
 2 3 4 5  =    1
 1 3 4 5  =    1
 1 2 4 5  =    1
 1 2 3 5  =    1
 1 2 3 4  =    1
\end{Soutput}
\end{Schunk}

Then we can use the {\tt grad()} function to calculate $d\omega$,
using the quotient law to express the derivatives analytically:

\begin{Schunk}
\begin{Sinput}
> df  <- function(x){
+     n <- length(x)
+     S <- sum(x^2)
+     grad(rep(c(1,-1),length=n)*(S^(n/2) - n*x^2*S^(n/2-1))/S^n
+     )
+ }
\end{Sinput}
\end{Schunk}

Thus

\begin{Schunk}
\begin{Sinput}
> df(1:5)
\end{Sinput}
\begin{Soutput}
An alternating linear map from V^1 to R with V=R^5:
             val
 1  =   4.05e-05
 2  =  -2.84e-05
 3  =   8.10e-06
 4  =   2.03e-05
 5  =  -5.67e-05
\end{Soutput}
\end{Schunk}

Now we can use the wedge product of the two parts (as per
equation~\ref{definition_of_d}) to show that the exterior derivative
of $\omega_n$, evaluated at a random point in $\mathbb{R}^n$, is
zero:

\begin{Schunk}
\begin{Sinput}
> x <- rnorm(9)
> print(df(x) ^ hat(x))  # should be zero
\end{Sinput}
\begin{Soutput}
An alternating linear map from V^9 to R with V=R^9:
                       val
 1 2 3 4 5 6 7 8 9  =    0
\end{Soutput}
\end{Schunk}

\subsection{Differential of the differential, $d^2=0$}

We can use the package to verify the celebrated fact that, for any
$k$-form $\phi$, $d\left(d\phi\right)=0$.  The first step is to define
scalar functions {\tt f1(), f2(), f3()}, all  $0$-forms:

\begin{Schunk}
\begin{Sinput}
> f1 <- function(w,x,y,z){x + y^3 + x*y*w*z}
> f2 <- function(w,x,y,z){w^2*x*y*z + sin(w) + w+z}
> f3 <- function(w,x,y,z){w*x*y*z + sin(x) + cos(w)}
\end{Sinput}
\end{Schunk}

Now we need to define elementary $1$-forms:

\begin{Schunk}
\begin{Sinput}
> dw <- as.kform(1)
> dx <- as.kform(2)
> dy <- as.kform(3)
> dz <- as.kform(4)
\end{Sinput}
\end{Schunk}

I will demonstrate the theorem by defining a $2$-form which is the sum
of three elementary two-forms, evaluated at a particular point in
$\mathbb{R}^4$:

\begin{Schunk}
\begin{Sinput}
> phi <-
+   (
+     +f1(1,2,3,4) ^ dw ^ dx
+     +f2(1,2,3,4) ^ dw ^ dy
+     +f3(1,2,3,4) ^ dy ^ dz
+   )
\end{Sinput}
\end{Schunk}

We could use slightly slicker R idiom by defining elementary forms
{\tt e1,e2,e3} and then defining {\tt phi} to be a linear sum, weighted with
$0$-forms given by the (scalar) functions {\tt f1,f2,f3}:

\begin{Schunk}
\begin{Sinput}
> e1 <- dw ^ dx
> e2 <- dw ^ dy
> e3 <- dy ^ dz
> phi <-
+   (
+     +f1(1,2,3,4) ^ e1
+     +f2(1,2,3,4) ^ e2
+     +f3(1,2,3,4) ^ e3
+   )
> phi
\end{Sinput}
\begin{Soutput}
An alternating linear map from V^2 to R with V=R^4:
              val
 1 2  =  53.00000
 1 3  =  29.84147
 3 4  =  25.44960
\end{Soutput}
\end{Schunk}

Now to evaluate first derivatives of {\tt f1()} etc at point
$(1,2,3,4)$, using {\tt Deriv()} from the Deriv package:

\begin{Schunk}
\begin{Sinput}
> library("Deriv")
> Df1 <- Deriv(f1)(1,2,3,4)
> Df2 <- Deriv(f2)(1,2,3,4)
> Df3 <- Deriv(f3)(1,2,3,4)
\end{Sinput}
\end{Schunk}

So {\tt Df1} etc are numeric vectors of length 4, for example:

\begin{Schunk}
\begin{Sinput}
> Df1
\end{Sinput}
\begin{Soutput}
 w  x  y  z 
24 13 35  6 
\end{Soutput}
\end{Schunk}

To calculate {\tt dphi}, or $d\phi$, we can use function {\tt grad()}:

\begin{Schunk}
\begin{Sinput}
> dphi <-
+   (
+     +grad(Df1) ^ e1
+     +grad(Df2) ^ e2
+     +grad(Df3) ^ e3
+   )
> dphi
\end{Sinput}
\begin{Soutput}
An alternating linear map from V^3 to R with V=R^4:
                val
 1 2 3  =  23.00000
 2 3 4  =  11.58385
 1 2 4  =   6.00000
 1 3 4  =  30.15853
\end{Soutput}
\end{Schunk}

Now work on the differential of the differential.  First evaluate
the Hessians (4x4 numeric matrices) at the same point:

\begin{Schunk}
\begin{Sinput}
> Hf1 <- matrix(Deriv(f1,nderiv=2)(1,2,3,4),4,4)
> Hf2 <- matrix(Deriv(f2,nderiv=2)(1,2,3,4),4,4)
> Hf3 <- matrix(Deriv(f3,nderiv=2)(1,2,3,4),4,4)
\end{Sinput}
\end{Schunk}

\begin{Schunk}
\begin{Sinput}
> rownames(Hf1) <- c("w","x","y","z")
> colnames(Hf1) <- c("w","x","y","z")
\end{Sinput}
\end{Schunk}

For example

\begin{Schunk}
\begin{Sinput}
> Hf1
\end{Sinput}
\begin{Soutput}
   w  x  y z
w  0 12  8 6
x 12  0  4 3
y  8  4 18 2
z  6  3  2 0
\end{Soutput}
\end{Schunk}

(note the matrix is symmetric; also note carefully the nonzero
diagonal term).  But $dd\phi$ is clearly zero as the Hessians are
symmetrical:

\begin{Schunk}
\begin{Sinput}
> ij <- expand.grid(seq_len(nrow(Hf1)),seq_len(ncol(Hf1)))
> ddphi <- # should be zero
+   (
+     +as.kform(ij,c(Hf1))
+     +as.kform(ij,c(Hf2))
+     +as.kform(ij,c(Hf3))
+   )
> ddphi
\end{Sinput}
\begin{Soutput}
The zero alternating linear map from V^2 to R with V=R^n:
empty sparse array with 2 columns
\end{Soutput}
\end{Schunk}

Above we see that $dd\phi$ is zero, as expected.

\section{Stokes's theorem}

In its most general form, Stokes's theorem states

\begin{equation}
\int_{\partial X}\phi=\int_Xd\phi
\end{equation}

where $X\subset\mathbb{R}^n$ is a compact oriented $(k+1)$-dimensional
manifold with boundary $\partial X$ and $\phi$ is a $k$-form defined
on a neighborhood of $X$.

We will verify Stokes, following 6.9.5 of \cite{hubbard2015} in which

\begin{equation}
\phi=
\left(x_1-x_2^2+x_3^3-\cdots\pm x_n^n\right)
\left(
\sum_{i=1}^n
dx_1\wedge\cdots\wedge\widehat{dx_i}\wedge\cdots\wedge dx_n
\right)
\end{equation}

(a hat indicates that a term is absent), and we wish to evaluate
$\int_{\partial C_a}\phi$ where $C_a$ is the cube $0\leq x_j\leq a,
1\leq j\leq n$.  Stokes tells us that this is equal to $\int_{C_a}
d\phi$, which is given by

\begin{equation}
d\phi = \left(
1+2x_2+\cdots + nx_n^{n-1}\right)
dx_1\wedge\cdots\wedge dx_n
\end{equation}

and so the volume integral is just

\begin{equation}
\sum_{j=1}^n
\int_{x_1=0}^a
\int_{x_2=0}^a
\cdots
\int_{x_i=0}^a
jx_j^{j-1}
dx_1 dx_2\ldots dx_n=
a^{n-1}\left(a+a^2+\cdots+a^n\right).
\end{equation}

Stokes's theorem, being trivial, is not amenable to direct numerical
verification but the package does allow slick creation of $\phi$:

\begin{Schunk}
\begin{Sinput}
> phi <- function(x){
+     n <- length(x)
+     sum(x^seq_len(n)*rep_len(c(1,-1),n)) * as.kform(t(apply(diag(n)<1,2,which)))
+ }
> phi(1:9)
\end{Sinput}
\begin{Soutput}
An alternating linear map from V^8 to R with V=R^9:
                           val
 1 2 3 4 6 7 8 9  =  371423053
 1 3 4 5 6 7 8 9  =  371423053
 1 2 3 4 5 6 7 9  =  371423053
 1 2 3 4 5 7 8 9  =  371423053
 1 2 4 5 6 7 8 9  =  371423053
 2 3 4 5 6 7 8 9  =  371423053
 1 2 3 4 5 6 8 9  =  371423053
 1 2 3 5 6 7 8 9  =  371423053
 1 2 3 4 5 6 7 8  =  371423053
\end{Soutput}
\end{Schunk}

Recall that {\tt phi} is a function that maps $\mathbb{R}^9$ to
8-forms.  Here we choose $\left(1,2,\ldots,9\right)\in\mathbb{R}^9$
and {\tt phi(1:9)} as shown above is the resulting 8-form.  Thus, if
we write $\phi_{1:9}$ for {\tt phi(1:9)} we would have
$\phi_{1:9}\colon\left(\mathbb{R}^9\right)^8\longrightarrow\mathbb{R}$,
with package idiom as follows:

\begin{Schunk}
\begin{Sinput}
> E <- matrix(runif(72),9,8)   # (R^9)^8
> as.function(phi(1:9))(E)
\end{Sinput}
\begin{Soutput}
[1] -26620528
\end{Soutput}
\end{Schunk}

Further, $d\phi$ is given by

\begin{Schunk}
\begin{Sinput}
> dphi <- function(x){
+     nn <- seq_along(x)
+     sum(nn*x^(nn-1)) * as.kform(seq_along(x))
+ }
> dphi(1:9)
\end{Sinput}
\begin{Soutput}
An alternating linear map from V^9 to R with V=R^9:
                             val
 1 2 3 4 5 6 7 8 9  =  405071317
\end{Soutput}
\end{Schunk}

Observe that {\tt dphi(1:9)} is a 9-form, with
$d\phi_{1:9}\colon\left(\mathbb{R}^9\right)^9\longrightarrow\mathbb{R}$.
Now consider Spivak's theorem 4.6 (page 82), which in this context
states that a 9-form is proportional to the determinant of the
$9\times 9$ matrix formed from its arguments, with constant of
proportionality equal to the form evaluated on the identity matrix
$I_9$ [formally and more generally, if $v_1,\ldots,v_n$ is a basis for
  $V$, $\omega\in\Lambda^n(V)$ and $w_i=\sum a_{ij}v_j$ then
  $\omega\left(w_1,\ldots,w_n\right) =
  \det\left(a_{ij}\right)\cdot\omega\left(v_1,\ldots v_n\right)$].
Numerically:

\begin{Schunk}
\begin{Sinput}
> f <- as.function(dphi(1:9))
> E <- matrix(runif(81),9,9)
> LHS <- f(E)
> RHS <- det(E)*f(diag(9))  
> c(LHS=LHS,RHS=RHS,diff=LHS-RHS) # LHS==RHS, according to Spivak's 4.6
\end{Sinput}
\begin{Soutput}
     LHS      RHS     diff 
-9850953 -9850953        0 
\end{Soutput}
\end{Schunk}

Above we see agreement to within numerical precision.

\section{Conclusions and further work}

The {\tt stokes} package furnishes functionality for working with
tensors, alternating forms, and related concepts from the exterior
calculus.  Theorems including Stokes's generalized integral theorem
were verified numerically.  Further work might include working with
Stokes's theorem expressed in Clifford algebra
formalism~\cite{hankin2022_clifford,hestenes1987}, following
Klausen~\cite{klausen2022}.

\bibliographystyle{apalike}
\bibliography{stokes}

\begin{thebibliography}{}

\bibitem[Hankin, 2022a]{hankin2022_clifford}
Hankin, R. K.~S. (2022a).
\newblock Clifford algebra in {R}.
\newblock \url{https://arxiv.org/abs/2209.13659}.

\bibitem[Hankin, 2022b]{hankin2022_disordR}
Hankin, R. K.~S. (2022b).
\newblock Disordered vectors in {R}: introducing the {{\tt disordR}} package.
\newblock \url{https://arxiv.org/abs/2210.03856}.

\bibitem[Hankin, 2022c]{hankin2022_spray}
Hankin, R. K.~S. (2022c).
\newblock Sparse arrays in {R}: the {{\tt spray}} package.
\newblock \url{https://arxiv.org/abs/2210.03856}.

\bibitem[Hestenes, 1987]{hestenes1987}
Hestenes, D. (1987).
\newblock {\em Clifford algebra to geometric calculus}.
\newblock Kluwer.

\bibitem[Hubbard and Hubbard, 2015]{hubbard2015}
Hubbard, J.~J. and Hubbard, B.~B. (2015).
\newblock {\em Vector calculus, linear algebra, and differential forms: a
  unified approach}.
\newblock Matrix Editions, fifth edition.

\bibitem[Klausen, 2022]{klausen2022}
Klausen, K.~O. (2022).
\newblock Visualizing {S}tokes' theorem with geometric algebra.
\newblock \url{https://arxiv.org/abs/2206.07177}.

\bibitem[{R Core Team}, 2022]{rcore2022}
{R Core Team} (2022).
\newblock {\em R: A Language and Environment for Statistical Computing}.
\newblock R Foundation for Statistical Computing, Vienna, Austria.

\bibitem[Spivak, 1965]{spivak1965}
Spivak, M. (1965).
\newblock {\em Calculus on manifolds}.
\newblock Addison-Wesley.

\end{thebibliography}

\end{document}